# Analysis and Design Specifications for Full-Duplex Radio Transceivers under RF Oscillator Phase-Noise with Arbitrary Spectral Shape

Ville Syrjälä, *Member, IEEE*, Koji Yamamoto, *Member, IEEE*, and Mikko Valkama, *Member, IEEE*

*Abstract*—In this paper, the effects of oscillator phase-noise with arbitrary spectral characteristics on self-interference cancellation capability of a full-duplex radio transceiver are addressed, and design considerations are given for oscillator designers for optimized PLL design for full-duplex radio application. The paper first gives a full-duplex transceiver model that inherently mitigates most of the phase-noise effect from the self-interference signal. The remaining effect of the phase noise is then analysed. Closed-form solutions for the self-interference power are then derived. In the simulations part, a practical phase-locked loop type oscillator is used, which is based on the arbitrary mask phase-noise model. Analytical derivations are verified with the simulations, and the self-interference cancellation performance is thoroughly studied with various parameters. Design considerations are finally given for oscillator design for full-duplex radio transceivers, with the help of tangible parameters of the phase-locked loop type oscillators.

## I. INTRODUCTION

FULL-DUPLEX radio is currently one of the hot research topics in the area of wireless communications. It enables the simultaneous use of the same radio frequency (RF) carrier for the transmission and reception. This offers benefits in terms of improved spectral efficiency and simpler frequency planning in communications networks. It also enables the transceiver to be aware of its signalling environment, because the transceiver is able to continuously receive the signals that

This work was supported in part by the Japan Society for the Promotion of Science (JSPS) (under the Postdoctoral Fellowship program and KAKENHI Grant number 25-03718 until September 2014), Emil Aaltonen Foundation, the Academy of Finland (under the projects 276378 "Cognitive Full-Duplex Radio Transceivers: Analysis and Mitigation of RF Impairments, and Practical Implementation" and 259915 "In-Band Full-Duplex MIMO Transmission: A Breakthrough to High-Speed Low-Latency Mobile Networks") and the Finnish Funding Agency for Technology and Innovation (under the project "Full-Duplex Cognitive Radio")

V. Syrjälä and M. Valkama are with the Department of Electronics and Communications Engineering (DECE), Tampere University of Technology, P.O. Box 553, 33101 Tampere, Finland. V. Syrjälä is also affiliated with the DECE.

K. Yamamoto is with the Graduate School of Informatics, Kyoto University, Yoshida-honmachi, Sakyo-ku, Kyoto-shi, 606-8501, Kyoto, Japan. V. Syrjälä was also affiliated with the Kyoto University until September 2014.

Corresponding author is Ville Syrjälä, e-mail: ville.syrjala@tut.fi

the other transceivers of the network transmit. This allows various benefits in the medium access control (MAC) layer design, and potentially minimizes the network signalling and delays, and therefore improves the overall throughput [1].

Practical implementation of full-duplex radio transceivers has not been considered possible until recently, because of the self-interference issue the technology faces. The self-interference stems from the coupling of the own transmitted signal to the receiver chain. This is a serious problem even in wireless local area networks (WLAN), where communications distances are relatively short, and the power differences between the transmitted and received signal powers are therefore also relatively smaller than, e.g., in macro cells of cellular mobile radio systems [1].

Recent research, e.g. in [1], [2], [3], [4], [5], has shown that full-duplex radio transceivers are implementable in laboratory conditions. It has also been shown in the literature that full-duplex radio transceivers are sensitive to various radio frequency (RF) impairments, especially to phase noise of the upconverting and downconverting oscillators at the transmitter and receiver parts, respectively [6], [7], [8], [9]. In [6], it was experimentally proved that the phase noise is indeed one of the most critical performance limiting factors in the self-interference cancellation of full-duplex radio transceivers. In [7], the phase noise effect was studied in more detailed manner, with analysis and simulations. A more realistic case of having the same oscillator feeding the oscillator signal to the transmitter and receiver parts of the transceiver was proposed, and analytically and numerically studied. Having the same oscillator signal used at the transmitter and receiver parts of the transceiver was shown to significantly decrease the effects of the phase noise. However, at the same time, phase noise was shown to be a great performance limiting factor even in the same oscillator case. The analytical studies thus far have all been based on the so-called free-running oscillator model [7], so it remains interesting to see the actual effects of the phase noise with more realistic oscillator models. This is the main topic of this article.

The main contributions of this paper are (*i*) the extension of the analysis of [7] for an *arbitrary oscillator* case (not limited to phase-locked loop (PLL) oscillators, even though the simulations part focuses on PLL case), namely to case where the phase noise of the oscillator has arbitrary spectrum shape,

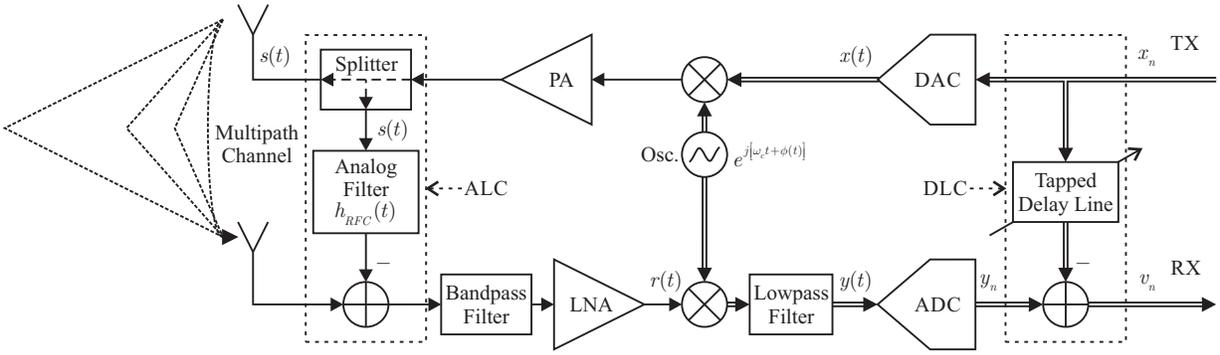

Fig. 1. Used full-duplex radio transceiver model with analog linear cancellation (ALC) and digital linear cancellation (DLC) blocks. Single lines denote real signals and double lines denote complex signals. TX and RX denote the transmitter and receiver parts, respectively. PA is power amplifier, LNA is low-noise amplifier, and DAC and ADC are digital-to-analog and analog-to-digital converters, respectively.

(*ii*) analysing the phase noise for the general type analog cancellation of self-interference (not only focusing on the direct coupling path of the channel), and (*iii*) the *design considerations for oscillator designers* to optimize PLL-oscillators for full-duplex transceiver application. Additional contributions of this paper are as follows: (*i*) simple and easy to understand baseband signal model and (*ii*) the effect of the PLL-oscillator generated phase noise on self-interference cancellation is extensively studied with various parameterizations in the simulations. Combined, these contributions are significant improvement over the previous work, because in practical transceivers, free-running oscillator phase-noise is not realistic. Therefore there is an urgent need for a more general oscillator model to be studied. Furthermore, there are infinite ways to design a PLL oscillator. Therefore, it is imperative that the oscillator designers are given design considerations how to optimize the PLL oscillator to be used in practical full-duplex radio transceivers. All the contributions are novel and not available in current literature.

This paper is structured as follows. The Section II gives the used full-duplex radio transceiver architecture. Also, full-duplex radio transceiver signal model is derived with phase noise taken into account using the transceiver model. In Section III, the power of the self-interference is analysed, and the used oscillator model is described and analysed. In Section IV, the used simulation environment and assumptions are described and extensive simulation results and analysis are given. Validity of the analysis is also verified with the simulations, and design considerations are given for oscillator designers. In the Section V, the work is concluded.

## II. Full-Duplex Radio Transceiver with Phase Noise

This section describes the used full-duplex radio transceiver architecture and provides a signal model used in the self-interference suppression analysis.

### A. Full-Duplex Transceiver Architecture

Fig. 1 depicts the proposed and used full-duplex radio transceiver architecture. The model consists of a direct conversion transmitter and receiver, with two feed-forward structures for self-interference cancellation. In the self-interference cancellation, first the upconverted and amplified signal is fed to the receiver through an analog filter. The analog filter mimics the multipath component of the self-interference channel [10]. The output of the component is then subtracted from the received signal, which effectively removes most of the self-interference propagated through the first multipath component of the self-interference channel. This is similar structure as used for analog cancellation in [1], while the more general structure with more complex analog filter is similar to the structure proposed in [10]. This process is from now on called analog linear cancellation (ALC). The ALC structure of [1] was selected because of its relatively low complexity and since the focus of this paper is on relatively small and energy efficient transceivers.

Then, the signal is received and the ALC is carried out, the signal is filtered, amplified, downconverted and once again filtered, as done also in conventional direct conversion receivers [11]. Now, after the sampling, digital domain self-interference cancellation is carried out. The process is here called digital linear cancellation (DLC). The digital signal from the transmitter is fed through a tapped delay line that tries to mimic the self-interference channel with the ALC taken into account. The output samples of the tapped delay line are then subtracted from the received sampled signal.

The proposed structure inherently mitigates the phase noise in the main component of the self-interference multipath channel by feeding the same oscillator signal to the upconverting oscillator at the transmitter and to the downconverting oscillator at the receiver.

As already discussed in [7], the delay difference between the oscillator signal fed to the transmitter and the one fed to the receiver causes only a small effect on the achievable self-interference cancellation level. However, the relatively small delay adds no complexity, and simplifies the modelling of the phase noise effect. The delay line length $l$ is naturally approximate, because of the air interface, but as was shown in [7], an error in order of few centimeters in the propagation estimate results in negligible error in the achievable self-interference cancellation. The delay line is relatively short, since the main multipath component (the direct coupling path) is expected to be short in a compact full-duplex radio transceiver.

## B. Baseband-Equivalent Signal Model with Phase Noise

The signal to be transmitted, $x(t)$, is generated from the digital samples $x_n$ by digital-to-analog converter. The signal is then upconverted and amplified resulting in a baseband-equivalent signal

$$s(t) = x(t)e^{j\phi(t)}. \tag{1}$$

Here, the amplification is assumed linear to separate the effect of phase noise that we are here interested in. In addition, all the amplification factors in this signal model are left out from the equations for the sake of simplifying and clarifying the presentation. This can be done since in the end all the signal powers are handled in relation to each other, so the actual amplification factors are not of interest. Notice also that the signal model is complex, even though, e.g., the RF signal is not complex. This is done also for the simplicity of the presentation, and can be done, because after the progressing through the full link, the final signals are the same anyway. In this way of modelling, the filters in the overall transceiver structure are already taken into account in the signal model, before they actually appear in the structure.

Now, at the receiver antenna, the self-interference signal has experienced the self-interference channel with baseband-equivalent impulse response $h(t)$. Also, the baseband-equivalent useful signal with additive noise and distortions and channel effects $u_{BB}(t)$ is added to the overall signal. Then, after the ALC, bandpass filtering and amplification (once again amplification and attenuation factors are left out), the baseband-equivalent signal can be written as

$$
\begin{aligned}
r(t) &= \big[h * s\big](t) - \big[h_{RFC} * s\big](t + \tau_e(t)) + u_{BB}(t) \\
&= \big[\tilde{h}_{ALC} * s\big](t)e^{j\omega_e\tau_e} + u_{BB}(t) \\
&= \big[h_{ALC} * s\big](t) + u_{BB}(t),
\end{aligned} \tag{2}
$$

where notation $\big[a * b\big](t)$ denotes the convolution between signals $a(t)$ and $b(t)$, $\tau_e$ is the delay difference between the direct coupling path of the channel and the tuneable delay for that path, and $h_{RFC}(t)$ is the baseband-equivalent impulse response of the tuneable attenuation and delay unit so that the delay of the direct coupling path of the channel has been taken out of it, $\tilde{h}_{ALC}(t)$ is the baseband-equivalent channel impulse response with the effect of the ALC taken into account, and finally $h_{ALC}(t)$ is $\tilde{h}_{ALC}(t)$ but also the rotation $e^{j\omega_e\tau_e}$ taken into account. Note that the timing is now given without any delays, even though strictly speaking the transmitted signal has already at this point experienced the delay. In the presentation of this paper, delays are always given in relation to the currently explained part of the self-interference link of the transceiver. Now after the downconversion and lowpass filtering, the baseband-equivalent signal can be written as

$$
\begin{aligned}
y(t) &= \big[h_{ALC} * s\big](t)e^{-j\phi(t+\tau_{\phi,e})} + u_{BB}(t) \\
&\approx \big[h_{ALC} * s\big](t)e^{-j\phi(t)} + u_{BB}(t).
\end{aligned} \tag{3}
$$

The small delay error $\tau_{\phi,e}$ (the delay difference between the actual propagation of the signal thorough the first multipath component of the channel and the excess delay of the oscillator signal from the oscillator to the downconverting mixer) causes negligible error in the signal, because it is very short delay [7], so it is omitted from the signal model in the second form of (3). The simulations show that even with half a meter antenna separation, this delay has only around 1 dB effect on the performance. With small integrated chips, on which this paper focuses on, the delay has no practical effect.

Next, the signal is sampled. By using the commonly used approach to model the channel with a tapped delay line in the digital domain, the sampled version of $y(t)$ can be written as

$$
\begin{aligned}
y_n &= y(nT_s) = \big[h_{n,ALC} * s_n\big]e^{-j\phi_n} + u_{n,BB} \\
&= \big[h_{n,ALC} * \big\{x_n e^{j\phi_n}\big\}\big]e^{-j\phi_n} + u_{n,BB} \\
&= \sum_{m=0}^{M} h_{m,ALC} x_{n-m} e^{j\big[\phi_{(n-m)} - \phi_n\big]} + u_{n,BB}.
\end{aligned} \tag{4}
$$

Here, the samples are taken at $t = nT_s : n \in \mathbb{Z}$, where $T_s$ is the sampling interval. In (4), $u_{n,BB} = u_{BB}(nT_s)$, $\phi_n = \phi(nT_s)$, and $h_{n,ALC}$ is the baseband-equivalent digital impulse response of the self-interference channel $h_{ALC}(t)$. The notation $a_n * b_n$ denotes the digital convolution between the sample streams $a_n$ and $b_n$, and $M$ denotes the maximum delay spread of the channel in samples.

The DLC is then performed. The signal samples $x_n$ that were transmitted at the transmitter are fed through a tapped delay line that tries to mimic the self-interference channel taking into account the ALC effect on the overall channel. The impulse response of the tapped delay line is therefore $\hat{h}_{n,ALC}$, namely the estimate of $h_{n,ALC}$. The samples at the output of the tapped delay line are then subtracted from the received signal samples $y_n$, resulting in a signal that can be written as

$$
\begin{aligned}
v_n &= \sum_{m=0}^{M} h_{m,ALC} x_{n-m} e^{j\big(\phi_{(n-m)} - \phi_n\big)} - \hat{h}_{m,ALC} x_{n-m} + u_{n,BB} \\
&= \sum_{m=0}^{M} \Big( h_{m,ALC} e^{j\big(\phi_{(n-m)} - \phi_n\big)} - \hat{h}_{m,ALC} \Big) x_{n-m} + u_{n,BB}.
\end{aligned} \tag{5}
$$

These are the received signal samples after the ALC and the DLC.

## C. Subcarrier-Wise Signal Model in OFDM Case

The phase noise is especially interesting in OFDM case since its effects on OFDM signals are severe and complex in nature [12], [13]. Therefore, in this paper, the phase noise effects are studied in more detail in OFDM case. Also, the fact that OFDM is utilized in most of the current and emerging communications standards motivates to focus this study on the OFDM waveforms. It is well-known that the phase noise has two distinct effects on OFDM signals [12], [13]. The first effect is so-called common phase error (CPE) which is a common rotation that all the subcarriers within one OFDM symbol experience. It is caused by the 0th frequency bin of the phase-noise complex-exponential. Since it is only a rotation, it cannot in practise be observed separately from the channel. It is therefore estimated automatically in the channel estimation, or if the channel is not estimated often enough, CPE is trivial to estimate in any case because of its simple nature. In the OFDM signal model of this paper, we model the CPE part of the phase-noise effect as part of the channel effect for simplicity. Also, the channel estimate used in DLC is assumed to have been obtained from the signal with CPE, because it is not realistic to assume an obtained channel estimate without the CPE already taken into account in it.

To study the subcarrier-wise effect of the phase noise, the signal $v_n$ is discrete Fourier transformed. The length $N$ discrete Fourier transform of the effective phase noise complex exponential, starting from the 0th sample, is (the $N$ is also the amount of subcarriers per OFDM symbol)

$$J_{k,m} = \frac{1}{N}\sum_{n=0}^{N-1} e^{j\left(\phi_{(n-m)} - \phi_n\right)} e^{-j2\pi kn/N} .$$ (6)

Notice now, that this frequency-domain phase-noise complex-exponential has value one as its 0th frequency bin, because the CPE is modelled as part of the channel $h_{m,ALC}$. Therefore, $\phi_{n-m} - \phi_n$ is the effective phase noise without the CPE part, namely approximately zero mean, which is anyway with any practical oscillator. Using (6), the discrete Fourier transformed $v_n$ in (5) can then be written as

$$
\begin{aligned}
V_k &= \mathrm{DFT}\left\{ v_n : n = 0, \dots, N-1 \right\} \\
&= \sum_{m=0}^{M}\left( h_{m,ALC} J_{k,m} - \hat{H}_{k,m,ALC} \right) * X_k e^{-j2\pi km/N} + U_{k,BB} ,
\end{aligned}
$$ (7)

where $\hat{H}_{k,m,ALC}$ is the discrete Fourier transform of the $m$th tap of the channel estimate. Since the tap of the channel estimate is constant, $\hat{H}_{k,m,ALC}$ has value $\hat{h}_{m,ALC}$ when $k = 0$ and value 0 for $k \neq 0$. $U_{k,BB}$ is the discrete Fourier transform of $u_{n,BB}$, and $X_k$ is the Fourier transform of the self-interference signal.

When writing the convolution in (7) as a sum, we arrive in form

$$
\begin{aligned}
V_k &= \sum_{m=0}^{M}\sum_{l=0}^{N-1}\left( h_{m,ALC} J_{k-l,m} - \hat{H}_{k-l,m,ALC} \right) X_l e^{-j2\pi lm/N} + U_{k,BB} \\
&= \sum_{m=0}^{M}\sum_{l=0}^{N-1}\left( h_{k-l,m,DLC} J_{k-l,m} \right) X_l e^{-j2\pi lm/N} + U_{k,BB} .
\end{aligned}
$$ (8)

Here, the second form is obtained because the CPE is modelled in the channel. Therefore, $h_{k,m,DLC}$ has value $h_{m,ALC} - \hat{h}_{m,ALC}$ when $k = 0$ and value $h_{m,ALC}$ when $k \neq 0$. $h_{k,m,DLC}$ is the channel effect after ALC and DLC.

## III. Subcarrier-Wise Self-Interference Power in OFDM under Arbitrary Oscillator

In this Section, the power of the subcarrier-wise self-interference is derived, the used oscillator model is presented, and then by using the oscillator model, the final closed form solutions for the self-interference power are derived.

### A. Power of Subcarrier-Wise Self-Interference

Now since we are interested in the self-interference power, we start from the signal in (8), but without the additive useful signal and noise term $V_{k,BB}$, namely without signal

$$I_k = \sum_{m=0}^{M}\sum_{l=0}^{N-1}\left( h_{k-l,m,DLC} J_{k-l,m} \right) X_l e^{-j2\pi lm/N} .$$ (9)

This is the subcarrier-wise self-interference. Its power can be written as

$$\mathrm{E}\left[ \left| I_k \right|^2 \right] = \mathrm{E}\left[ \left| \sum_{m=0}^{M}\sum_{l=0}^{N-1}\left( h_{k-l,m,DLC} J_{k-l,m} \right) X_l e^{-j2\pi lm/N} \right|^2 \right].$$ (10)

Here $\mathrm{E}\left[ \cdot \right]$ is the statistical expectation operator and $\left| \cdot \right|$ is the absolute value operator. Now, with reasonable assumption that $\forall k : X_k$ are independent of each other and zero mean, and by using the widely used Bello's wide-sense stationary uncorrelated scattering (WSSUS) model [14] for the channel ( $\forall m : h_{k,m,DLC}$ are independent of each other and zero mean), we can rewrite (10) as

$$
\begin{aligned}
E\left[ \left| I_k \right|^2 \right] &= \mathrm{E}\left[ \left| \sum_{l=0}^{N-1} X_l \sum_{m=0}^{M} h_{k-l,m,DLC} J_{k-l,m} e^{-j2\pi lm/N} \right|^2 \right] \\
&= \sum_{l=0}^{N-1} \mathrm{E}\left[ \left| X_l \right|^2 \right] \sum_{m=0}^{M} \mathrm{E}\left[ \left| h_{k-l,m,DLC} J_{k-l,m} e^{-j2\pi lm/N} \right|^2 \right] \\
&= \sum_{l=0}^{N-1}\sum_{m=0}^{M} \sigma_l^2 g_{k-l,m,DLC} \, \mathrm{E}\left[ \left| J_{k-l,m} \right|^2 \right].
\end{aligned}
$$ (11)

Here, in the last form $\sigma_l^2$ is $\mathrm{E}\left[ \left| X_l \right|^2 \right]$ and $g_{k,m,DLC}$ is $\left| h_{k,m,DLC} \right|^2$, namely the gain of the effective channel after ALC and DLC. To obtain the closed form solution for the self-interference power, we need to derive $\mathrm{E}\left[ \left| J_{k,m} \right|^2 \right]$, and for that we need to fix an oscillator model.

### B. Generalized Oscillator Model with Arbitrary Phase Noise Spectral Shape

Instead of the more academic free-running oscillator model, the oscillator model in this paper is based on the generic spectral mask model proposed in [15] and used also, e.g., in [16]. With the model, we can, e.g., model phase noise of realistic PLL based oscillators [17]. The idea in the model is to generate white Gaussian noise, transfer it to the frequency domain with discrete Fourier transform, multiply the frequency domain phase-noise with a desired spectral mask of the phase noise, and transfer the result back to the time domain with inverse discrete Fourier transform . Naturally this generates the phase noise in blocks whose length comes from the length of the discrete Fourier transform. This is sufficient for our purposes, since we are only interested in analysing an average phase-noise effect within an OFDM symbol on self-interference cancellation.

Our goal now is to derive $\mathrm{E}\left[\left|J_{k,m}\right|^2\right]$ for the above described oscillator model. This is done by deriving $\mathrm{E}\left[\left|J_{k,m}\right|^2\right]$ into a form that can be expressed as a function of a phase-noise mask. Let us start by applying well-known small-phase approximation to $J_{k,m}$. The approximation is very accurate, because for practical PLL oscillators that are useable in full-duplex in the first place, practical offsets from the zero phase are very small (the maximum is in order of 0.1 rad). This value projects into maximum errors in order of 0.5 % in the real part of the phase noise complex exponential. With the approximation, $J_{k,m}$ can be written as

$$J_{k,m} \approx 1 + \frac{j}{N}\sum_{n=0}^{N-1}\left[\phi_{(n-m)} - \phi_n\right]e^{-j2\pi kn/N} \,. \tag{12}$$

Now, if we denote the discrete Fourier transform of the phase noise $\phi_n$ as $\Phi_k$, we can rewrite (12) into form

$$J_{k,m} \approx \begin{cases} 1 + j\left(\Phi_k e^{-j2\pi km/N} - \Phi_k\right), & k=0 \\ j\left(\Phi_k e^{-j2\pi km/N} - \Phi_k\right), & k\neq0 \end{cases} \\ = \begin{cases} 1 + jO_{k,m}, & k=0 \\ jO_{k,m}, & k\neq0 \end{cases} \tag{13}$$

where

$$O_{k,m} = \Phi_k e^{-j2\pi km/N} - \Phi_k \,. \tag{14}$$

Now,

$$\mathrm{E}\left[\left|J_{k,m}\right|^2\right] \approx \begin{cases} \mathrm{E}\left[1 + O_{k,m}O_{k,m}^* + jO_{k,m} - jO_{k,m}^*\right], & k=0 \\ \mathrm{E}\left[O_{k,m}O_{k,m}^*\right], & k\neq0 \end{cases} \\ = \begin{cases} 1 + \mathrm{E}\left[O_{k,m}O_{k,m}^*\right], & k=0 \\ \mathrm{E}\left[O_{k,m}O_{k,m}^*\right], & k\neq0 \end{cases} \,. \tag{15}$$

Here, $(\cdot)^*$ denotes the complex conjugation, and the second form is obtained due to the linearity of the statistical expectation operator, and because the phase noise is zero mean (it is spectral masked zero-mean white Gaussian noise) and therefore the statistical expectation of $O_{0,m}$ is unity. In (15), the power of $O_{k,m}$ can be written into form (analysed separately to simplify the presentation)

$$\mathrm{E}\left[O_{k,m}O_{k,m}^*\right] = \mathrm{E}\left[2\Phi_k\Phi_k^* - \Phi_k\Phi_k^*e^{-j2\pi km/N} - \Phi_k\Phi_k^*e^{j2\pi km/N}\right] \\ = 2\,\mathrm{E}\left[\Phi_k\Phi_k^*\right] - 2\cos\left(2\pi km/N\right)\mathrm{E}\left[\Phi_k\Phi_k^*\right]. \tag{16}$$

Therefore, $\mathrm{E}\left[\left|J_{k,m}\right|^2\right]$ finally arrives into form

$$\mathrm{E}\left[\left|J_{k,m}\right|^2\right] \approx \begin{cases} 1 + 2\sigma_{k,PN}^2\left[1 - \cos\left(\dfrac{2\pi km}{N}\right)\right], & k=0 \\ 2\sigma_{k,PN}^2\left[1 - \cos\left(\dfrac{2\pi km}{N}\right)\right], & k\neq0 \end{cases}. \tag{17}$$

Here, $\sigma_{k,PN}^2 = \mathrm{E}\left[\Phi_k\Phi_k^*\right]$ and it is the average frequency-domain power of the phase noise calculated for subcarrier index $k$. It can therefore be used as a spectrum mask, so the initial goal of this subsection of deriving $\mathrm{E}\left[\left|J_{k,m}\right|^2\right]$ into a form that it is possible to be expressed with a help of a spectrum mask is now achieved.

Notice that in this subsection, $\mathrm{E}\left[\left|J_{k,m}\right|^2\right]$ was also calculated for $k=0$ for completeness of the analysis of the phase-noise power. However, in the signal model $\mathrm{E}\left[\left|J_{0,m}\right|^2\right]$ is unity as already explained during the derivation of the signal model.

### C. Closed-Form Solution for Power of Subcarrier-Wise Self-Interference

Now we can use $\mathrm{E}\left[\left|J_{k,m}\right|^2\right]$ derived in the previous subsection, and rewrite (11) with the help of approximation in (17) as

$$E\left[\left|I_k\right|^2\right] \approx \sum_{m=0}^{M}\left\{\sigma_k^2 g_{0,m,DLC} \\ + \sum_{\substack{l=0 \\ l\neq k}}^{N-1}\sigma_l^2 g_{k-l,m,DLC}\, 2\sigma_{k-l,PN}^2\left[1 - \cos\left(\dfrac{2\pi km}{N}\right)\right]\right\}. \tag{18}$$

Since the CPE error part in (18) is separate, it is convenient to return to original notation for the channel used prior the second form of (8), and rewrite (18) into its final form

$$E\left[\left|I_k\right|^2\right] \approx \sum_{m=0}^{M}\left\{\sigma_k^2\left|h_{m,ALC} - \hat{h}_{m,ALC}\right|^2 \\ + 2\sum_{\substack{l=0 \\ l\neq k}}^{N-1}\sigma_l^2\left|h_{m,ALC}\right|^2\sigma_{k-l,PN}^2\left[1 - \cos\left(\dfrac{2\pi km}{N}\right)\right]\right\}. \tag{19}$$

This is an easy and efficient tool to calculate the self-interference power in closed-form for oscillator phase-noise with arbitrary spectral shape. Notice that the variables $\sigma_{k,PN}^2$ characterize the oscillator phase noise characteristics, as the phase noise power per frequency bin.

### IV. SIMULATOR, SIMULATION RESULTS AND ANALYSIS

This section gives a description of the used simulator, verification of the derived closed-form analysis results and extensive simulations analysis in various cases. Also, the design considerations for PLL type oscillator design for full-duplex radio transceivers are discussed while the results are analysed. In addition, the increased performance of the phase noise mitigation of the proposed new structure is numerically compared to the structure proposed in [7], alongside with the delay estimation error effect in the variable delay for the oscillator signal.

### A. Simulator

#### 1) Simulation Routine

The simulator first generates 16QAM subcarrier data symbols. These are then OFDM modulated using 1024 subcarriers per OFDM symbol, so that 300 on the both sides of the middle subcarrier are active and the others are null subcarriers. Used sampling frequency is 15.36 MHz, resulting in 15 kHz subcarrier spacing. Then, cyclic prefix of 63 samples is added. These parameters were selected because the system now resembles the 3G Partnership Project (3GPP) Long Term Evolution (LTE) downlink signal with 10 MHz effective bandwidth [18], [19]. Therefore, the simulation parameters are near to ones used in a modern communications standard. At this point, the transmitter phase noise is modelled into the signal with the model explained in the next subsection, after which the coupling channel from transmitter to receiver is modelled. The default used channel has power delay profile of $0$ dB, $-65$ dB, $-70$ dB and $-75$ dB with delays 0, 1, 2 and 4, respectively ($M = 4$). This resembles closely the channel measured for full-duplex relays in [20], but it is modified to better suit for the full-duplex radio transceivers. It is relatively short channel, but that is expected because it is a channel between transmitter and receiver antennas that are very close to each other. The channel is normalized so that the response of the main multipath component is on average unity, so the antenna separation is modelled separately. The antenna separation is modelled so that the main multipath component is attenuated enough that a desired level of antenna separation is attained. The antenna separation is assumed to affect only the main multipath component, because it is very likely that by increasing the separation between antennas in a way that it is possible in a compact radio transceiver, the other multipath components cannot be much affected. In these simulation trials the antenna separation is fixed at 30 dB. Making effectively the default power delay profile of the used channel as $-30$ dB, $-65$ dB, $-70$ dB and $-75$ dB with delays 0, 1, 2 and 4, respectively, since the main coupling path is so dominating. This corresponds the physical power delay profile of the channel.

At the receiver, ALC is modelled. It is modelled so that additive white Gaussian noise is added to the used channel tap so that a desired level of ALC is achieved. This means that the ALC only removes the self-interference from the first channel tap for simplicity (notice that the closed-form solution does not have this restriction.) Amount of ALC denotes the suppression of the whole signal including the other multipath components, so the main multipath component may need more suppression than the desired amount of ALC is. In this study, the default ALC level is fixed at 30 dB. Exception is the perfect ALC case, where no white Gaussian noise is added to the known channel tap in the ALC process. Notice that maximum achievable ALC is at around 33.5 dB for the case of the default channel, if only first tap of the channel is used in

ALC. After the ALC, the receiver phase noise is modelled. The same phase noise process is used as was used in the transmitter part. Finally, the DLC is modelled. In the DLC, the estimation error is modelled by adding white Gaussian noise to the individual channel taps. The level of the added white Gaussian noise is determined by the amount of DLC desired. In this study, the desired DLC is set at 70 dB (this value was selected so that the total suppression is around what is required by modern communications standards [7]). Once again, the exception is the case of perfect DLC, when no additive noise is added. After this, the signal power is evaluated to compute the attained self-interference cancellation level.

Notice that the channel estimation errors are modelled so that the additive white Gaussian noise is added to the actual channel taps, even though the phase noise is present. This is because the channel estimation is done in frequency domain as is usually done for OFDM signals. The CPE part of the phase noise error was assumed to be estimated with the channel, which is realistic assumption because its effect is only a phase shift in the symbol constellation. The ICI part of the phase noise error on the other hand causes the energy of all subcarriers to spread on the full-band of the OFDM symbols. The error is only nearly Gaussian [21] for general free-running oscillators. In the case of the proposed transceiver structure, the inherent phase noise mitigation heavily whitens the remaining phase noise, making it practically white noise. It is therefore realistic assumption that the estimation error in the channel estimation of the channel impulse response is white Gaussian noise.

#### 2) Short Description of Used PLL Oscillator Model

The PLL oscillator model used in the simulations is the so-called charge-pump PLL (CHPLL) model proposed in [22] and described also in more detail in [17] and [23]. In the model, the CHPLL oscillator phase-noise is generated with a spectral mask of the phase noise. Therefore, it fits perfectly for the study of this paper, since in this paper we also use the phase-noise mask to characterize the oscillator phase-noise. In the model, CHPLL phase-noise spectral characteristics are generated based on so-called single-side-band phase-noise measurement results from the crystal oscillator (CO) and the voltage-controlled oscillator (VCO) used in the CHPLL oscillator. For CO, there is no flicker noise region (slope $-20$ dB/dec ), so only a measurement from the thermal noise dominated region (slope $-30$ dB/dec ) is required [23]. In this paper, for the CO, $-160$ dBc/Hz performance at 1 MHz offset from the oscillation frequency is assumed. This is very typical value [17]. For the VCO, two measurements are considered. One from the flicker noise dominated region (typically from 1 kHz to 100 kHz offset from the oscillation frequency). In this paper the measurement is always assumed to be taken at 1 kHz offset from the oscillation frequency and the actual measurement result is denoted by $L_f$. On the other hand, the thermal noise measurement for the VCO (typically taken at frequencies between 1 MHz to 3 MHz) is assumed to be taken at 1 MHz offset from the oscillation frequency and the result is denoted by $L_w$.

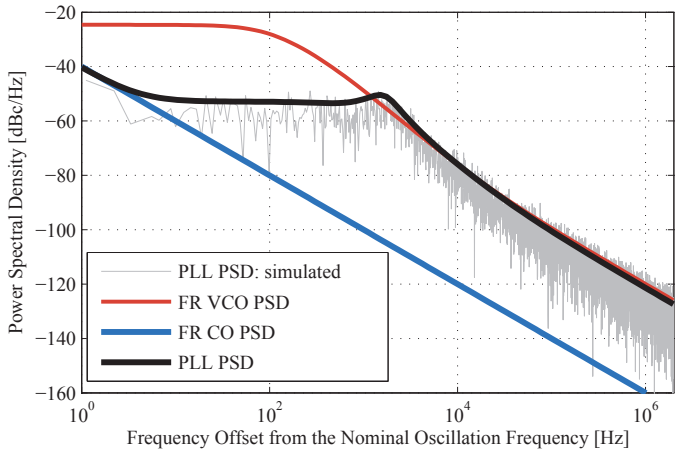

Fig. 2. An example spectrum illustration of the spectrum generated by the used oscillator model.

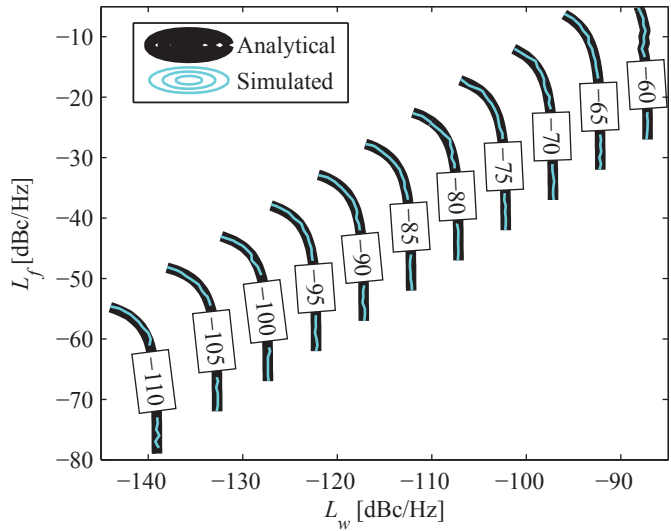

Fig. 3. Contour plot giving the achievable self-interference suppression levels as a function of oscillator parameters $L_f$ and $L_w$, when ALC is 30 dB, DLC with ideal channel knowledge is used, and antenna separation is 30 dB. The curves are cut because of parameter limitations.

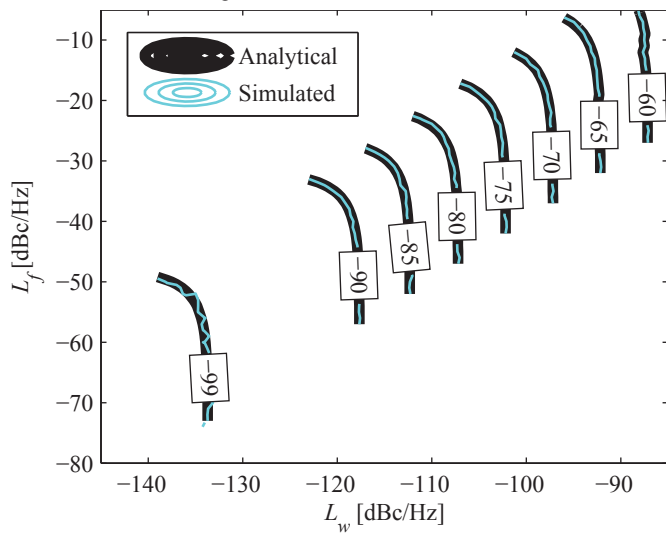

Fig. 4. Contour plot giving the achievable self-interference suppression levels as a function of oscillator parameters $L_f$ and $L_w$, when ALC is 30 dB, DLC is 70 dB, and antenna separation is 30 dB. The last contour level is set to -99 dB because -100 dB is only achievable for ideal oscillator (when $L_f$ and $L_w$ would be $-\infty$). The curves are cut because of parameter limitations.

Notice that the parameter values depend on each other. Since the measurements of $L_f$ and $L_w$ are taken 3 decades apart from each other (at 1 kHz and at 1 MHz), when the $L_f$ is in the beginning of the flicker noise region and $L_w$ is in the beginning of the thermal noise region, we have limitation $L_f + 60 \text{ dB} \leq L_w$. Another limitation comes in a case where the $L_f$ is taken at the point just between the flicker noise and thermal noise regions. Then the limitation is $L_f + 90 \text{ dB} \geq L_w$. Therefore, the combined limitation for $L_w$ given $L_f$ is $L_f + 60 \text{ dB} \leq L_w \leq L_f + 90 \text{ dB}$. Naturally, both of the extreme values are rather unrealistic, but theoretically possible.

An example spectrum of the phase noise process generated with the used oscillator model, with VCO parameters $L_f = -76 \text{ dBc/Hz}$ at 10 kHz offset and $L_w = -120 \text{ dBc/Hz}$ at 1 MHz offset, is illustrated in Fig. 2.

*3) Motivation to Used Full-Duplex Parameters*

The fixed value of 30 dB of antenna separation is around typical values reported in the literature, as is the ALC of 30 dB for the studied transceiver architecture [1], [2], [3]. The 70 dB DLC on the other hand is relatively optimistic value [1], [2], [3]. However, in the previous literature it is clear that the attainable DLC was heavily limited by, e.g., phase noise [6], [7], and here the DLC denotes the attainable DLC level without any transceiver impairments. Therefore, the set level 70 dB is quite well motivated. Also, the case of ideal DLC is very interesting, because in that case we can entirely focus on studying the limits of self-interference cancellation imposed by the phase noise.

### B. Simulation Results, Analysis and Design Considerations

In the simulations, the attainable self-interference cancellation is always studied in different cases. In all the given simulations, the attainable self-interference cancellation is the total self-interference cancellation with the phase noise after the ALC and DLC are used. Therefore, in the case of ALC of 30 dB and DLC of 70 dB, the achievable self-interference cancellation without phase noise is always 100 dB (ALC+DLC), and in the ideal DLC case the corresponding cancellation is $\infty$ dB. For ideal ALC and 70 dB DLC case, the perfect result is around 103.5 dB. This is because the default channel limits the maximum attainable ALC (see the definition of the ALC level in the first subsection of this section) to around 33.5 dB.

The simulation results as a function of the oscillator parameters $L_f$ and $L_w$ are given in Fig. 3, Fig. 4, Fig. 5 and Fig. 6. Notice that the limitation of useable values of $L_f$ and $L_w$ explained in previous subsection is also visible in the figures. In Fig. 3 and Fig. 4, contour plots of achievable self-interference cancellation levels are given as a function of both of the parameters $L_f$ and $L_w$. In Fig. 3, perfect channel knowledge is used in the DLC. With reasonable oscillator parameter values $L_f$ and $L_w$, around 110 dB level of self-interference cancellation (ALC+DLC) is attainable at best. This is already a good result, but very high-quality VCO would be needed in the PLL to obtain such results. Already parameter levels of $L_f = -60 \text{ dBc/Hz}$ and $L_w = -120$ dBc/Hz result into a high quality oscillator, and the attainable self-interference cancellation is then at around 93 dB. This level is also a good achieved value, but might not be enough for many of the communications standards. As an example, in LTE uplink, the maximum transmitted signal power is 23 dBm [18], so the signal is at $-7$ dBm after the 30 dB antenna separation. The sensitivity level can in worst case be at around $-110$ dBm [19], so the required self-interference cancellation can be even somewhere between 103 and 110 dB. Therefore, it is indeed very interesting to optimize the oscillator design as carefully as possible. In Fig. 4, the results are given when the DLC is limited to 70 dB. It can be seen that if the DLC is limited, then still extremely high-quality oscillator is required to get 99 dB of self-interference cancellation, even though the level restricted by the channel estimation errors is at 100 dB.

In Fig. 5 and Fig. 6, we have the achievable self-interference levels as a function of $L_f$ and $L_w$, respectively. These figures show how great effect individual parameter $L_f$ or $L_w$ has on the achievable self-interference cancellation level with DLC of 70 dB or with ideal channel knowledge. In Fig. 5, $L_w = -120$ dBc/Hz, $L_f \geq L_w - 90$ dB and $L_f \leq L_w - 60$ dB. We can see that improving $L_f$ from its minimum value to the maximum value gives a change of around 6 dB in the achievable self-interference cancellation level. The maximum gain is not much considering the huge change in the oscillator quality in terms of $L_f$. However, already the change from the maximum $L_f$ of $-30$ dBc/Hz to around $-40$ dBc/Hz provides considerable share of the maximum available gain. In this region therefore, the optimization of the parameter $L_f$ is very beneficial. In Fig. 6, $L_f = -50$ dBc/Hz and $L_f + 90$ dB $\leq L_w \leq L_f + 60$ dB . Now, we can see that all over the whole theoretical design region of $L_w$, lowering its value provides almost constant, and significant, gains to the achievable self-interference cancellation, expect when we get to around $L_w = -130$ dBc/Hz level, after which the gain begins to get smaller. At this point $L_f$ starts to limit the performance as demonstrated by Fig. 3 and Fig. 4.

Overall, from the parameters $L_f$ and $L_w$, $L_w$ has clearly the dominating effect on the achievable self-interference cancellation levels. However, at lower $L_w$ levels $L_f$ indeed starts to limit the performance. It is therefore very challenging optimization task to find a good complexity level in PLL-oscillator design and circuit implementation costs while keeping a good full-duplex radio performance. Optimizing $L_f$ can relax the requirements for $L_f$ around 5 dB or so, even with a slight adjustment in it.

In Fig. 7, the self-interference cancellation level is analysed as a function of DLC with fixed oscillator parameters $L_f = -50$ dBc/Hz and $L_w = -120$ dBc/Hz , assuming either ALC with perfect channel knowledge (results into ALC of around 33.5 dB with the used channel) or with ALC of 30 dB. The behaviour of both of the curves is similar. The perfect ALC gives a bit better performance, and the performance difference disappears shortly after the phase noise starts to limit the performance of the DLC after DLC of around 55 dB. So the effect of the phase noise is visible already at relatively low levels of DLC.

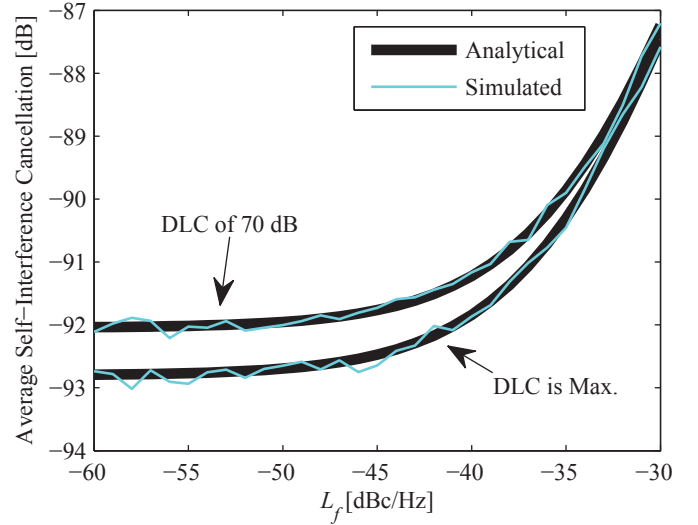

Fig. 5. Average self-interference cancellation given as a function of $L_f$, when DLC is either 70 dB or DLC is made with perfect channel knowledge (Max.). Antenna separation is 30 dB and $L_w = -120$ dBc/Hz .

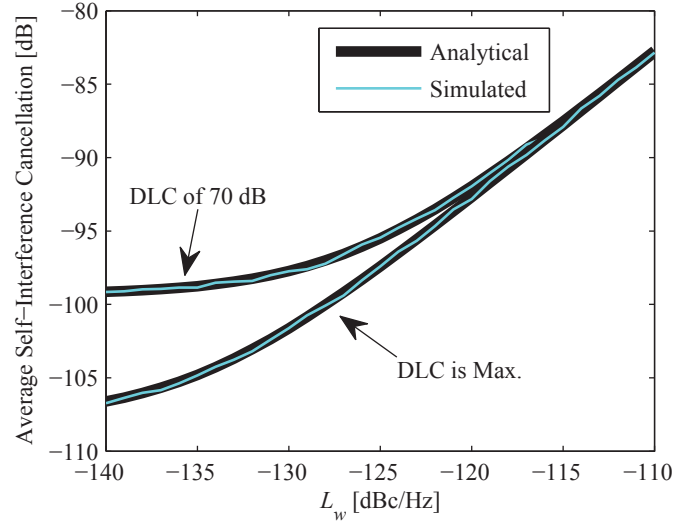

Fig. 6. Average self-interference cancellation given as a function of $L_w$, when DLC is either 70 dB or DLC is made with perfect channel knowledge (Max.). Antenna separation is 30 dB and $L_f = -50$ dBc/Hz .

In all simulated scenarios, the results given by the derived closed-form analytical formula are compared to results given by the simulations. In all of the cases, the analytical formulas give almost perfectly the same results as the simulator, giving strong indication that the derived closed form expressions and used signal models are indeed valid.

## C. Impact of Varying the Coupling Channel Characteristics

In this subsection, the default coupling channel is not used, but instead the effect of different type of channels on the oscillator design is studied. For simplicity and to study the effect of the key properties of the channel, namely the attenuation and delay, certain elementary channels are used. In this case, elementary channel means that we study two-tap channel, so that in the first case (Fig. 8) the second tap experiences one sample delay and its attenuation is varied, and in the second case (Fig. 9) the second tap experiences different delay and the attenuation is kept at constant value of $-65$ dB (resulting in channel with power delay profile of 0 dB and $-65$ dB for zero-delay and the varied delay, respectively.) In both of the cases, the ALC is ideal for the direct coupling path, so that the effect of the interesting tap (attenuated or delayed second tap) is clearly distinguished from the trivial direct component. DLC is also ideal (but without the knowledge of the phase noise), so that the only visible effect is the effect of the phase noise that is the only effect of interest in this paper.

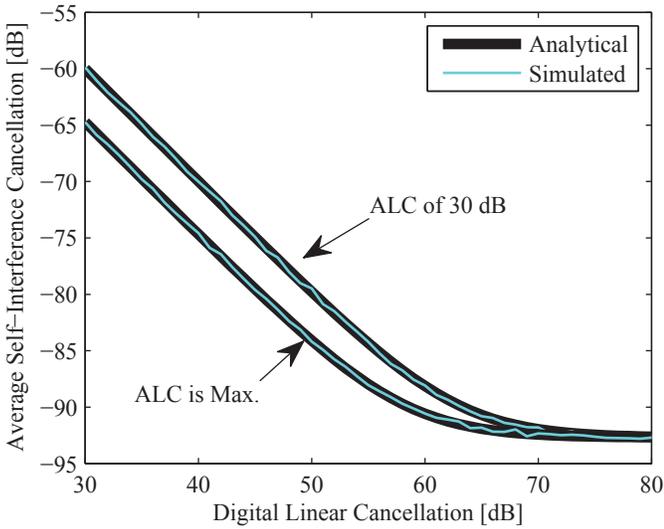

Fig. 7. Average self-interference cancellation given as a function of digital cancellation level, when ALC is either 30 dB or ALC is made with perfect channel knowledge (Max.). Antenna separation is 30 dB and $L_f = -50$ dBc/Hz and $L_w = -120$ dBc/Hz .

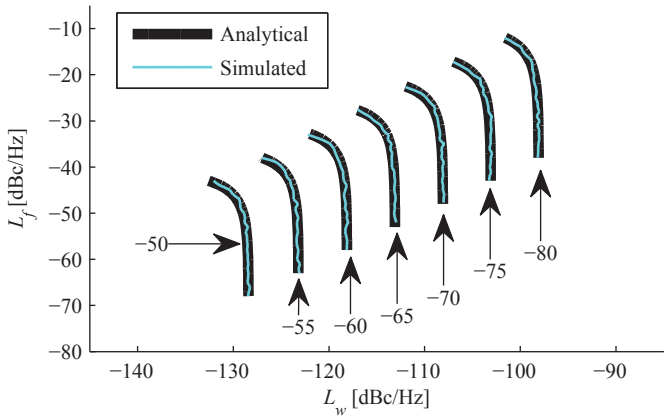

Fig. 8. Contour plot giving the achievable self-interference suppression levels as a function of oscillator parameters $L_f$ and $L_w$, when the attenuation of the second tap of the two-tap channel is varied. The varied amount is denoted by the arrows with attenuation in dB. For simplicity of the presentation only contour plots of $-90$ dB are given for each attenuation value. The curves are cut because of parameter limitations.

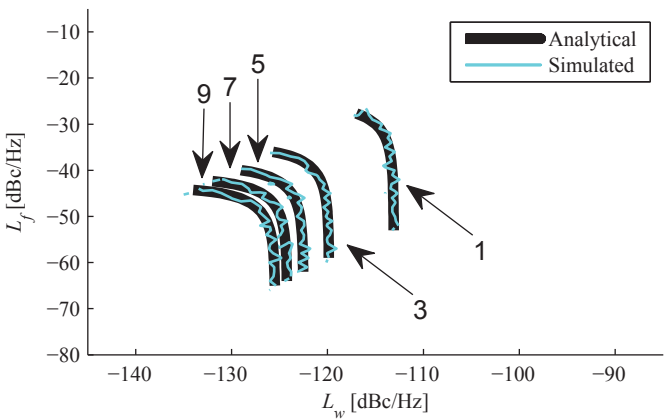

Fig. 9. Contour plot giving the achievable self-interference suppression levels as a function of oscillator parameters $L_f$ and $L_w$, when the delay of the second tap of the two-tap channel is varied. The varied amount is denoted by the arrows with the delay in samples. For simplicity of the presentation only contour plots of $-90$ dB are given for each delay value. The curves are cut because of parameter limitations.

From Fig. 8 we can see that changing the attenuation of the second tap has only easy-to-understand effect on the oscillator design. It just linearly increases the amount of self-interference. The shape of the curves are a bit different from the curves in Fig. 3, because the channel has indeed effect on the optimal oscillator design since it shapes the effective phase noise from the DLC perspective. From Fig. 9 we can see that by changing the delay, huge change in the shape of the contours are visible. This is because changing the delay effectively filters the phase noise, seen by DLC, and the filtering is different with different delays. We can see that the oscillator design is the easiest with short delays. This is natural, because higher delay directly results in higher difference between the phase noise of the upconverting and downconverting oscillators. With longer delays, even small decrease of $L_f$ allows us to relax the requirements for $L_w$ significantly, when the design is in the highest acceptable region of $L_f$. This is because the longer delay components provide weaker lowpass effect in the effective phase noise, and since $L_f$ is measured at relatively low-frequencies, it becomes more and more significant factor. At shorter delays, $L_w$ is clearly more a critical factor, which is natural since the lowpass effect is much stronger.

This study demonstrated that also the coupling channel plays a critical role in the oscillator design for full-duplex radio transceivers. The derived formulas therefore provide also from this point-of-view very effective way to help in oscillator design specifications for various different coupling channel scenarios. Furthermore, even though a full-duplex device architecture with separate transmit and receive antennas was assumed in the developments in this paper, the results are also applicable in shared-antenna based devices [10] deploying, e.g., a circulator to separate the transmit and receive chains. In this case, there are two dominant coupling components, namely direct leakage through the circulator and the reflection from the antenna. Hence, the results with the elementary two-path channel can be applied, to devise appropriate oscillator design specifications.

## V. Conclusion

Phase noise is one of the critical performance limiting factors in self-interference suppression in full-duplex radio transceivers, especially when targeting high digital cancellation levels. In this paper, a structure that inherently mitigates the self-interference on the first multipath component of the self-interference channel was proposed. Then, the signal model for the remaining signal was derived, and a closed-form expression of the power of the self-interference after analog linear cancellation and digital linear cancellation was derived, in a generic case with arbitrary spectral shape of the oscillator phase-noise. This is seen to be a big step compared to earlier reported analysis in literature, which build primarily on simple free-running oscillator model, and thus do not model the realistic behaviour of real oscillators used in state-of-the-art transceivers. The derived closed-form expressions can be used as a tool by oscillator designers to optimize the oscillator designs and specifications for full-duplex radio transceivers. In the simulations, the analysis results were verified to be highly accurate. It was also concluded that the phase noise levels were still significant even though the phase noise of the main multipath component was completely mitigated in the analog RF cancellation which motivates for the development of more advanced wideband RF cancellation solutions, also from the oscillator phase noise specifications perspective.

This paper also provides oscillator designers some design considerations and specifications for the design of PLL type oscillators. It was also concluded that even with very high-quality PLL oscillator, the phase noise can be huge issue if the full-duplex radio transceivers is used in modern communications standards. It was also shown, that the coupling channel multipath profile plays significant role in oscillator design for full-duplex radio transceivers. The derived tool is therefore expected to be very useful for the future transceiver component designers.

## REFERENCES

[1] M. Jain, J. Choi, T. Kim, D. Bharadia, K. Srinivasan, S. Seth, P. Levis, S. Katti, and P. Sinda, "Practical, real-time, full duplex wireless," in *Proc. International Conference on Mobile Computing and Networking 2011* (*MobiCom'11*), Las Vegas, NV, USA, Sep. 2011.

[2] J. Choi, M. Jain, K. Srinivasan, P. Levis, and S. Katti, "Achieving single channel, full duplex wireless communication," in *Proc. International Conference on Mobile Computing and Networking 2010* (*MobiCom'10*), Chicago, IL, USA, Sep. 2010.

[3] M. Duarte and A. Sabharwal, "Full-duplex wireless communications using off-the-shelf radios: Feasibility and first results," in *Proc. Asilomar Conference on Signals, Systems, and Computers*, Pacific Grove, CA, USA, Nov. 2010.

[4] E. Everett, A. Sahai, and A. Sabharwal, "Passive self-interference suppression for full-duplex infrastructure nodes," *IEEE Transactions on Wireless Communications*, Vol. 13, No. 2, Feb. 2014.

[5] A. Sabharwal, P. Schniter, D. Guo, D. Bliss, S. Rangarajan, and R. Wichman, "In-band full-duplex wireless: Challenges and opportunities," *IEEE Journal on Selected Areas in Communications*, Vol. 32, No. 10, Oct. 2014.

[6] A. Sahai, G. Patel, C. Dick, and A. Sabharwa, "On the impact of phase noise on active cancellation in wireless full-duplex," *IEEE Transactions on Vehicular Technology*, Vol. 62, No. 9, Nov. 2013.

[7] V. Syrjälä, M. Valkama, L. Anttila, T. Riihonen, and D. Korpi, "Analysis of oscillator phase-noise effects on self-interference cancellation in full-duplex OFDM radios," in *IEEE Transactions on Wireless Communications*, Vol. 13, No. 6, June 2014.

[8] D. Korpi, T. Riihonen, V. Syrjälä, L. Anttila, M. Valkama, and R. Wichman, "Full-duplex transceiver system calculations: Analysis of ADC and linearity challenges," *IEEE Transactions on Wireless Communications*, Vol. 13, No. 7, Jul. 2014.

[9] D. Korpi, L. Anttila, V. Syrjälä, and M. Valkama, "Widely-linear digital self-interference cancellation in direct-conversion full-duplex transceiver," *IEEE Journal on Selected Areas in Communications*, Vol. 32, No. 10, Oct. 2014.

[10] D. Bharadia, E. McMilin, and S. Katti, "Full duplex radios," in SIGCOMM'13, Aug. 2013.

[11] B. Razavi, "Design considerations for direct-conversion receivers," *IEEE Transactions on Circuits and Systems-II: Analog and Digital Signal Processing*, Vol. 44, No. 6, June 1997.

[12] P. Robertson and S. Kaiser, "Analysis of the effects of phase-noise in orthogonal frequency division multiplex (OFDM) systems," in *Proc. IEEE International Conference on Communications*, June 1995, pp. 1652-1657, Vol. 3.

[13] L. Tomba, "On the effect of Wiener phase noise in OFDM systems," *IEEE Transactions on Communications*, Vol. 46, No. 5, pp. 580-583, May 1998.

[14] P. Bello, "Characterization of randomly time-variant linear channels," *IEEE Transactions on Communications Systems*, Vol. 11, No. 4, pp. 360-393, December 1963.

[15] Y. Ferdi, A. Taleb-Ahmed, and M. Lakehal, "Efficient generation of 1/f^beta using signal modeling techniques," *IEEE Transactions on Circuits and Systems-I: Regular Papers*, Vol. 55, No. 6, pp. 1704-1710, July 2008.

[16] V. Syrjälä, M. Valkama, Y. Zou, N. N. Tchamov, and J. Rinne, "On OFDM link performance under receiver phase noise with arbitrary spectral shape," in *Proc. IEEE Wireless Communications & Networking Conference 2011* (*WCNC'11*), Cancun, Quintana-Roo, Mexico, March 2011.

[17] N. N. Tchamov, V. Syrjälä, J. Rinne, M. Valkama, Y. Zou, and M. Renfors, "System and circuit-level optimization of PLL design for DVB-T/H receivers," *Analog Integrated Circuits and Signal Processing Journal*, doi: 10.1007/s10470-011-9823-2, January 2012.

[18] 3GPP Technical Specification, TS 36.101 v. 8.20.0, E-UTRA UE Radio Transmission and Reception, Release 8, 2013.

[19] H: Holma and A. Toskala, "UTRAN Long-Term Evolution, " in *WCDMA for UMTS: HSPA Evolution and LTE*, 5th edition, John Wiley & Sons Ltd., 2010.

[20] K. Haneda, E. Kahra, S. Wyne, C. icheln, and P. Vainikainen, "Measurement of loop-back interference channels for outdoor-to-indoor full duplex radio relays," in *Proc. European Conference on Antennas and Propagation* (*EuCAP'10*), Barcelona, Spain, April 2010.

[21] T. Schenk, R. van der Hofstad, E. Fledderus, and P. Smulders, "Distribution of the ICI term in phase noise impaired OFDM systems," in *IEEE Transactions on Wireless Communications*, vol. 6, no. 4, April 2007.

[22] N. N. Tchamov, J. Rinne, V. Syrjälä, M. Valkama, Y. Zou, and M. Renfors, "VCO phase noise trade-offs in PLL design for DVB-T/H receivers," in *Proc. IEEE International Conference on Electronics, Circuits and Systems 2009* (*ICECS'09*), Yasmine Hammamet, Tunisia, December 2009.

[23] V. Syrjälä, *Analysis and Mitigation of Oscillator Impairments in Modern Receiver Architectures*, D.Sc. (Tech.) Dissertation, Tampere University of Technology, June 2012.

Placeholder

**Ville Syrjälä** (S'09–M'12) was born in Lapua, Finland, in 1982. He received the M.Sc. (Tech.) degree in 2007 and D.Sc. (Tech.) degree in 2012 in communications engineering (CS/EE) from Tampere University of Technology (TUT), Finland.

He was working as a Research Fellow with the Department of Electronics and Communications Engineering at TUT, Finland, until September 2013, and in Graduate School of Informatics, Kyoto University, Japan, as JSPS Postdoctoral Fellow until September 2014. Currently, he is working as An Academy Postdoctoral Researcher at TUT, Finland. His general research interests are in full-duplex radio technology, communications signal processing, transceiver impairments, signal processing algorithms for flexible radios, transceiver architectures, direct sampling radios, and multicarrier modulation techniques.

Placeholder

**Koji Yamamoto** (S'04–M'06) received the B.E degree in electrical and electronic engineering from Kyoto University in 2002, and the M.E. and Ph.D. degrees in informatics from Kyoto University in 2004 and 2005, respectively. From 2004-2005 he was a research fellow of the Japan Society for the Promotion of Science (JSPS). Since 2005, he has been with the Graduate School of Informatics, Kyoto University, where he is currently an Associate Professor. From 2008 to 2009, he was a visiting researcher at Wireless@KTH, Royal Institute of Technology (KTH) in Sweden. His research interests include game theory, spectrum sharin, and cooperative multi-hop networks. He received the PIMRC 2004 Best Student Paper Award in 2004, the Ericsson Young Scientist Award in 2006, and the Young Researcher's Award from the IEICE in 2008.

Placeholder

**Mikko Valkama** (S'00–M'02) was born in Pirkkala, Finland, on November 27, 1975. He received the M.Sc. and Ph.D. degrees (both with honours) in electrical engineering (EE) from Tampere University of Technology (TUT), Finland, in 2000 and 2001, respectively. In 2002 he received the Best Ph.D. Thesis award by the Finnish Academy of Science and Letters for his dissertation entitled "Advanced I/Q signal processing for wideband receivers: Models and algorithms".

In 2003, he was working as a visiting researcher with the Communications Systems and Signal Processing Institute at SDSU, San Diego, CA. Currently, he is a Full Professor and Department Vice Head at the Department of Electronics and Communications Engineering at TUT, Finland. He has been involved in organizing conferences, like the IEEE SPAWC'07 (Publications Chair) held in Helsinki, Finland. His general research interests include communications signal processing, estimation and detection techniques, signal processing algorithms for software defined flexible radios, full-duplex radio technology, cognitive radio, digital transmission techniques such as different variants of multicarrier modulation methods and OFDM, radio localization methods, and radio resource management for ad-hoc and mobile networks.